\documentclass{sig-alternate}
\begin{document}

% Page heads
% \markboth{V. H. P. Louzada et al.}{Measuring the Robustness of Boolean Networks}
%
\conferenceinfo{ACM-BCB '12,} {October 7-10, 2012, Orlando, FL, USA}
\CopyrightYear{2012}
\crdata{978-1-4503-1670-5/12/10}
%\CopyrightYear{2012}
%\crdata{978-1-4503-1670-5/12/10} 
%\conferenceinfo{ACM-BCB'12}{October 7-10, 2012, Orlando, FL, USA}
%
% Title portion
\title{A Monte Carlo Approach to Measure the Robustness of Boolean Networks}
\numberofauthors{3}
\author{
\alignauthor
Vitor H. P. Louzada \\
\affaddr{Computational Physics, IfB, ETH-Honggerberg, Switzerland }
\email{louzada@ethz.ch}
\alignauthor
Fabr\'{i}cio M. Lopes \\
\affaddr{Federal University of Technology - Paran\'a, Brazil}
\email{fabricio@utfpr.edu.br}
\alignauthor
Ronaldo F. Hashimoto \\
\affaddr{Institute of Mathematics and Statistics of the University of S\~ao
Paulo,S\~ao Paulo, Brazil}\\
\email{ronaldo@ime.usp.br}
}
\maketitle
\begin{abstract}
Emergence of robustness in biological networks is a par\-a\-mount feature of
evolving organisms, but a study of this property \textit{in vivo}, for any 
level of representation such as Genetic, Metabolic, or Neuronal Networks, 
is a very hard challenge. In the case of Genetic Networks, mathematical 
models have been used in this context to provide insights on their
robustness, but even in relatively simple formulations, such as 
Boolean Networks~(BN), it might not be feasible to compute some measures 
for large system sizes. We describe in this work a Monte Carlo approach 
to calculate the size of the largest basin of attraction of a BN, which
is intrinsically associated with its robustness, that can be used 
regardless the network size. We show the  stability of our method through 
finite-size analysis and validate it with a full search on small networks.
\end{abstract}
%\vspace{-1mm}
% A category with the (minimum) three required fields
%A category including the fourth, optional field follows...
%\vspace{-1mm}
%\vspace{-1mm}
%\acmformat{Louzada, V. H. P., Lopes, F. M., Hashimoto, R. F.  2012. A Monte Carlo Approach to Measure the Robustness of Boolean Networks.}
%\maketitle

\section{Introduction}
Robustness in biological organisms is one of the major characteristics which
contributes to their survival in the environment, maintaining its functions in
face of external and internal perturbations \cite{Kitano04,Kitano07}. Despite
its importance, the complete understanding of the mechanisms behind this
underlying property is still not possible, as \textit{in vivo} studies are a
very hard challenge due to knowledge limitations. In this context, mathematical
abstractions of the interactions that constitute an organism are powerful tools
to provide explanations about biological robustness, especially regarding the
robustness present in some biological networks~\cite{Li06,Serra04}. 

In general, it is expected that phenotype robustness is a consequence of a
specific set of rules or patterns in the interrelationship among the genes of an
organism~\cite{Stearns02}. Hence, theoretical models of Gene Regulatory Networks
might be able to provide insights on their robustness. As a relatively simple
and rich model, the Boolean Network (BN) model, which consider genes as on-off
switches, is often successfully applied when details about gene-gene
interactions are
absent~\cite{lopes2008a,Albert08,lopes2009a, Higa11,Kauffman69,Kauffman71,Kauffman93,Lopes11,lopes2011b}. 
Among several key features, attractors (which captures the gene expression patterns
that are periodically visited) and the size of their basins (which are made up
of all the expression patterns that conduct to this attractors) are measures
that can reveal important characteristics of the underlying biological
network~\cite{Li04,Louzada11}. More information regarding prospects and
limitations of this paradigm can be found in the review written by
Bornholdt~\cite{Bornholdt08}.

The simplification of the Boolean formalism does not solve completely the
problem of building a mathematical definition of robustness~\cite{Kitano07}. A
lot of measures have been proposed following different interpretations of the
concept, such as Derrida curves~\cite{Derrida86}, identification of
Intrinsically Multivariate Prediction Genes \cite{Martins08}, and the size of
the largest basin of attraction. In this work, we consider robustness as the
ability of executing the same activity despite random fluctuations in a limited
number of genes. Hence, a network with a large basin of attraction might be
considered robust, and the the size of the largest basin of attraction
($\Lambda$) a good estimation for its robustness.

The ability to measure the largest basin of attraction of a BN in a fast and
reliable way will cause a huge impact in the robustness characterization of the
most of the available Genetic Networks, which can easily be composed of more
than 6000 genes. However, measuring $\Lambda$ imposes its own challenge, as an
exhaustive search has an exponential complexity in the number of genes, making
it impractical for large network sizes. In Brun et al.~\cite{Brun05} an
estimation of $\Lambda$ is calculated from the steady state distribution of a
Probabilistic Boolean Network, but also with an exponential complexity in the
number of genes (nodes). This work proposes a Monte Carlo approach to calculate
$\Lambda$ that can be used regardless the network size. It is showed that the
proposed methodology is reliable through finite-size analysis. The obtained
results were validated by using a full search on small networks.

\section{Boolean Network}
\label{sec:mm}
A Boolean Network (BN) is defined by a set of $n$ Boolean variables $X=\{x_1,
x_2, \ldots, x_n\}$
and a set of $n$ Boolean functions $F=\{f_1, f_2, \ldots, f_n\}$. In our
context, each variable $x_i$ represents a \emph{gene}, and it can assume only
two possible values: $0$ (OFF) or $1$ (ON). 
The value of gene $x_i$ at time $t+1$ is obtained from the values of a set of
predictor genes $G_i=\{x_{j_1(i)},x_{j_2(i)},\ldots,x_{j_{k_i}(i)}\}\subseteq X$
at time $t$ through the Boolean function \mbox{$f_i:\{0,1\}^{k_i}\to\{0,1\}$}
that belongs to $F$. 

The number of predictors of $x_i \in X$ is called the \emph{in-degree} of $x_i$,
and the \emph{out-degree} of $x_i$ is defined as the number of genes in $X$ for
which $x_i$ is one of their predictors. We consider in this work that all genes
are updated synchronously by the functions in $F$ assigned to them.

A \emph{state} of a BN at time $t$ is a binary vector $s(t) =
(x_1(t),\ldots,x_n(t))$, representing the value of all Boolean variables in the
network, roughly describing the genetic activity of the organism (all genes) at
a certain time. In this way, the number of all possible states is $2^n$. Keeping
the Boolean function that regulates each gene fixed, and without considering any
random fluctuation in the value of the genes, a BN is a deterministic
formulation. 

The dynamics of a BN can be represented by a directed network, called
\emph{state transition diagram}, in which its nodes correspond to the states of
the BN and an arc from one node to another  corresponds to a state transition
between them. A set of states that is periodically visited is called an
\emph{attractor} of the BN, and all states that eventually lead to this
attractor (and including all states in the attractor) are called the \emph{basin
of attraction}. In this work, consider the number of states in the basin of
attraction as its \emph{size}. Finally, define $\Lambda$ as the size of the
largest basin of attraction divided by the total number of states.

\section{Method Description}

In our method, we assume that the largest basin of attraction is also the
easiest to identify, in the sense that a randomly chosen state is more likely to
be part of it. Hence, the probability that a randomly chosen state is the
largest basin of attraction is equal to the fraction $\Lambda$ of states in this
basin.

To estimate $\Lambda$, a Monte Carlo approach is proposed for large networks
since a full search is impracticable. We draw a number $Z$ of states and for
each of them the attractor is identified. The most frequent attractor has
$Z^{*}$ random states that conduct to it, hence we can assume that the ratio
$Z^{*}/Z$ is a good estimator for $\Lambda$, i.e., the size of the largest basin
divided by the total number of states. 

The number of states in an attractor, as well as the average number of time
steps that a state in the basin of attraction takes to reach its attractor, were
numerically estimated as the function $y(n) \cong 0.003n^{3.3294}$. This in
agreement with results provided by Kauffman~\cite{Kauffman93}, who considered
the attractor size as a polynomial function of the number of nodes. Hence, by
considering a randomly chosen network state, we can assume that the number of
time steps that it will take to reach all the states in its attractor (i.e., to
reach one state in the attractor and, from that state, visit all states in the
attractor), is, for most of the cases, $2y(n)$.  This upper bound implies that
the complexity of our algorithm is $2y(n)Z$ in the average case. 

\section{Results and Discussion}

The proposed methodology was executed considering different network sizes: $n$
from $10$ to $26$. Each gene regulatory network is generated as a random graph
(Erd\H{o}s-R\'{e}nyi (ER) directed graph, with in-degree fixed), characterized
by a Poisson distribution of out-degrees. For comparison, we perform a full
search of the largest basin of attraction on the smaller gene network sizes,
$n\leq 20$.  For each gene regulatory network size, the value of $\Lambda$ is
the average of 100 randomly generated networks. 

\begin{figure}[ht!]
\centerline{\includegraphics[width=0.95\linewidth]{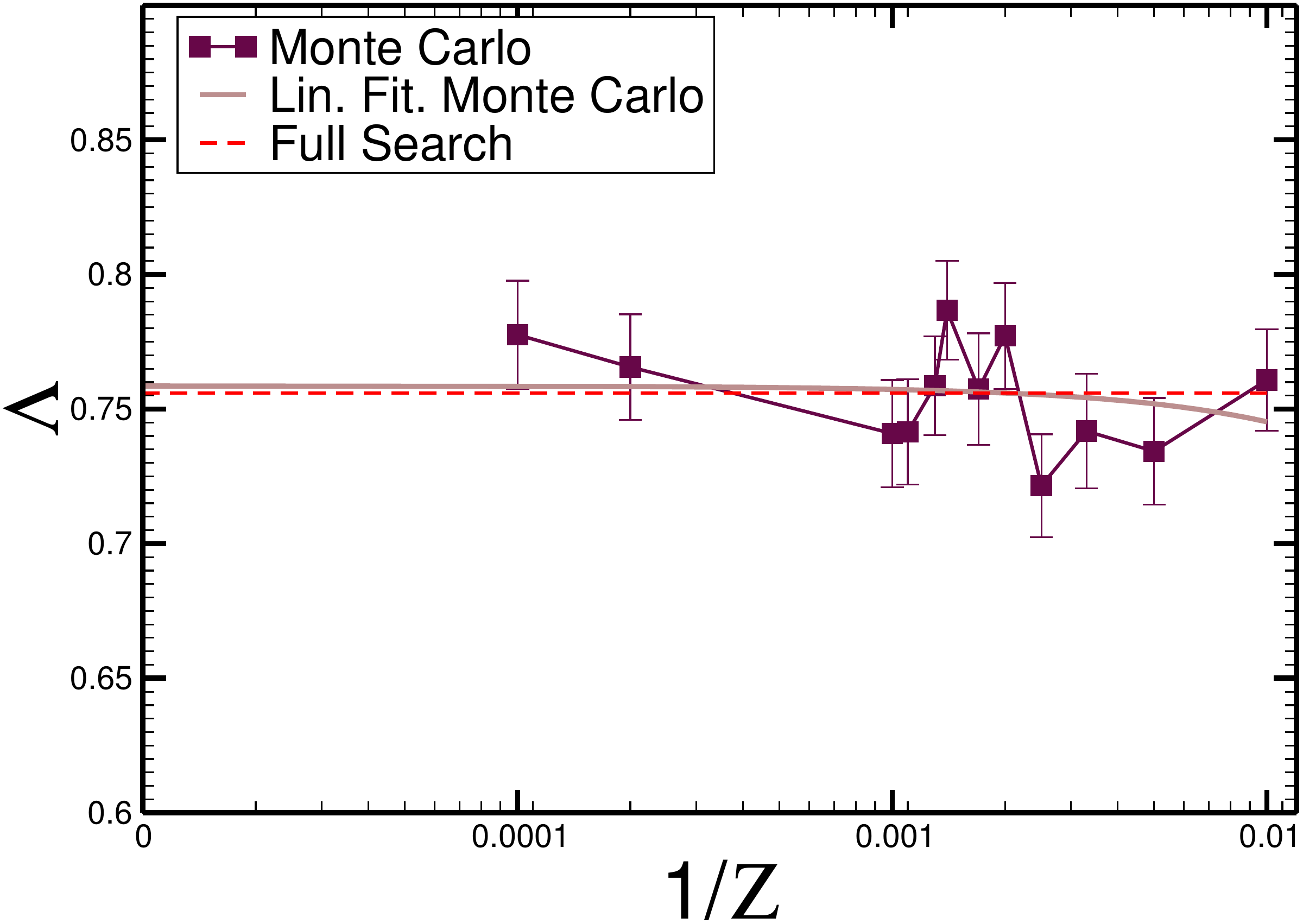}}
\caption{\textbf{Effect of the number of initial random states $Z$ on the
estimated largest basin of attraction $\Lambda$}. Purple squares represent our
Monte Carlo approach, with the light purple line as a Linear Fitting of the
data. In the limit of infinite $Z$ states, the extrapolation of our method is
very similar to the full search estimation (red line). The value of $\Lambda$ is
an average over 100 BNs of size $n=20$.}
\label{fig::ZversusL}
\end{figure}

\begin{figure}[hb!]
\centerline{\includegraphics[width=0.95\linewidth]{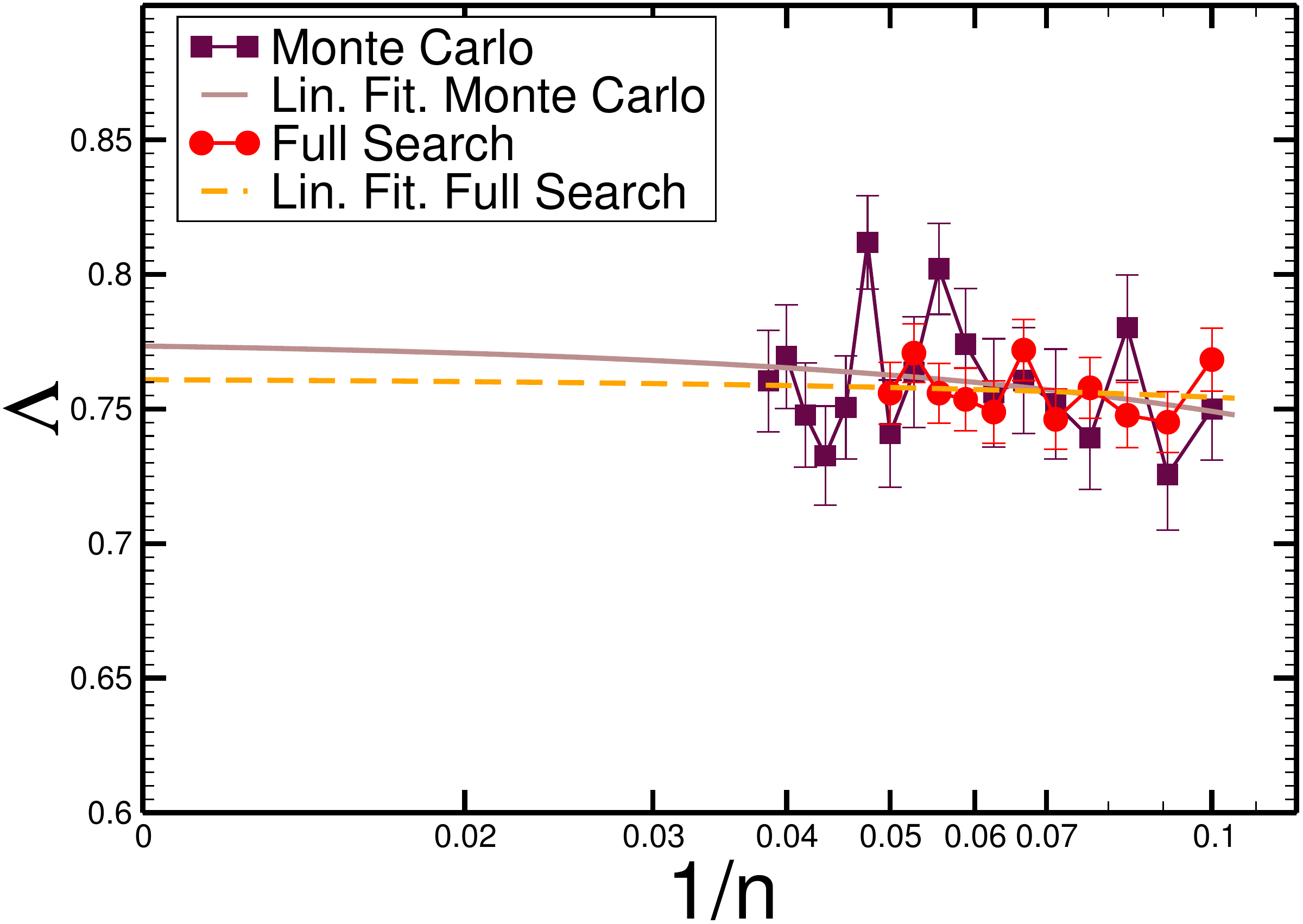}}
\caption{\textbf{Effect of the size $n$ of the BN on the estimated largest basin
of attraction $\Lambda$}. Purple/red squares represent our Monte Carlo
approach/Full search, with the light purple/orange line as a Linear Fitting of
the data. In the limit of infinite size, the extrapolation of both methods are
very similar. The value of $\Lambda$ is an average over 100 BNs. The Monte Carlo
method was executed with $Z=1000$.}
\label{fig::NversusL}
\end{figure}

In our proposal, the number of initial random states $Z$ is a very important
parameter to be determined, 
as if $Z \ll 2^n$ our method is much less costly than the exponential complexity
of the full search. 
By computing the value of $\Lambda$ against $1/Z$ in Figure~\ref{fig::ZversusL},
we show that:  a value for $Z$, as small as $1000$, already provides a good
estimation for a network of $2^{20}$ states,  and that the full search
estimation, $\Lambda=0.76$, is recovered for a sufficient large value of $Z$.

For a fixed value of $Z$ at $1000$, we measure $\Lambda$ for different network
sizes in Figure~\ref{fig::NversusL}, which also presents the value of $\Lambda$
calculated through a full search for small network sizes. Our proposal estimates
$\Lambda$ within the interval $[0.73,0.81]$, in accordance with the full search,
despite the fact that $Z=1000$ is much smaller than $2^n$, for $n>15$. Both
methods also point to a similar value of $\Lambda = 0.77$ in networks of
infinite size. 

\section{Conclusion}
\label{sec:conc}

We describe in this work a fast and reliable strategy to measure the largest
basin of attraction of a BN.
The proposed method has a polynomial complexity on the number of states and the
obtained results is in full agreement with full search results for small
networks, in the limit of an infinite number of genes. 
Besides that, our work is a proof-of-concept that Monte Carlo estimation might
be successfully applied in measures related to Boolean Networks.

We hope with this method to provide a powerful tool to future studies about
robustness in biological organisms. 
From this, it is possible to compare, for instance, the robustness of the Gene
Regulatory Networks of the yeast with a random network of the same size, 
roughly with $6000$ genes, possibly identifying features that enhance the size
of the largest basin of attraction. 

There are rooms for improvement in our method as well. As shown by
Linch~\cite{Lynch93,Lynch95}, the average size of the attractors $y(n)$ is
superpolinomial in the number of nodes. Hence, for large values of $n$, the
complexity of our method could be better estimated if the function $y(n)$ were
precisely described. Besides that, the choice of $Z$ could be improved, as a
better estimation could be made considering $Z$ as an increasing function of
$n$.

It is known that some biological networks have different topologies, such as
scale-free~\cite{stuart2003,albert2005} and small-world~\cite{ws1998}. Another
possible direction would be to apply the proposed methodology in larger networks
with different topologies. Also, topological characteristics of the state
transition diagram could be used to improve the efficiency of our method, for
instance considering more tractable BNs where the input of each Boolean function
is equal to one.

% Acknowledgments
\section{Acknowledgments}
The authors would like to thank the Brazilian agencies CAPES, FAPESP,
Microsoft-Research and Conselho Na\-ci\-o\-nal de Pesquisa (CNPq) for the
financial support.

% Bibliography
%\bibliographystyle{acmsmall}
\bibliographystyle{abbrv}
\bibliography{stabilityBN}

\end{document}